\documentclass[11pt]{article}
\usepackage{graphicx}
\usepackage{amssymb}

\newcommand{\BABARPubYear}    {02}

\newcommand{\BABARConfNumber} {012}
\newcommand{\SLACPubNumber} {9282}

\input pubboard/babarsym

\long\def\inst#1{\par\nobreak\kern 4pt\nobreak
    {\it #1}\par\vskip 10pt plus 3pt minus 3pt}

\begin{document}
{\pagestyle{empty}

\begin{flushright}
\babar-CONF-\BABARPubYear/\BABARConfNumber \\
SLAC-PUB-\SLACPubNumber \\
July 2002 \\
\end{flushright}

\par\vskip 5cm

\begin{center}
\Large \bf Measurement of the Inclusive Electron Spectrum in Charmless
Semileptonic $B$ Decays Near the Kinematic Endpoint
\end{center}
\bigskip

\begin{center}
\large The \babar\ Collaboration\\
\mbox{ }\\
July 24, 2002
\end{center}
\bigskip \bigskip

\begin{center}
\large \bf Abstract
\end{center}
A preliminary result on the study of the inclusive electron
spectrum in $B\to X_u e \nu$ decays above the kinematic
limit for the dominant $B\to X_c e \nu$  transitions is presented.
This study is performed at the \pep2 \abf, where $B$ meson pairs are produced
in the decay of the \FourS resonance.
For the electron momentum range of $2.3 - 2.6$~\gevc\
in the \FourS\ rest frame, the partial branching ratio is measured to be
$\Delta \textrm{\BR}(B\rightarrow X_ue\nu)=
(0.152\pm 0.014\pm 0.014)\cdot10^{-3}$.

\vfill
\begin{center}
Contributed to the 31$^{st}$ International Conference on High Energy Physics,\\ 
7/24---7/31/2002, Amsterdam, The Netherlands
\end{center}

\vspace{1.0cm}
\begin{center}
{\em Stanford Linear Accelerator Center, Stanford University, 
Stanford, CA 94309} \\ \vspace{0.1cm}\hrule\vspace{0.1cm}
Work supported in part by Department of Energy contract DE-AC03-76SF00515.
\end{center}

\newpage
} 

\begin{center}
\small

The \babar\ Collaboration,
\bigskip

B.~Aubert,
D.~Boutigny,
J.-M.~Gaillard,
A.~Hicheur,
Y.~Karyotakis,
J.~P.~Lees,
P.~Robbe,
V.~Tisserand,
A.~Zghiche
\inst{Laboratoire de Physique des Particules, F-74941 Annecy-le-Vieux, France }
A.~Palano,
A.~Pompili
\inst{Universit\`a di Bari, Dipartimento di Fisica and INFN, I-70126 Bari, Italy }
% ...

\end{center}\newpage

\section{Introduction} 

The principal physics goal of the \babar\ experiment is to establish 
$CP$-violation 
in $B$ mesons and to test whether the observed effects are
consistent with the prediction of the Standard Model.  In this model 
$CP$ violating effects are predicted from the CKM matrix of the
couplings of the charged weak current to quarks.  A precise determination of
the matrix element $|V_{ub}|$ will place constraints on the unitarity of the
CKM matrix and thus the consistency with the minimal Standard Model.

The extraction of $|V_{ub}|$ is a challenge, both theoretically and 
experimentally.  
While at the parton level weak interactions can be reliably calculated, 
meson decays   
depend on the $b$ quark mass and its motion inside the $B$ meson. 
Theoretical
calculations of the semileptonic decay rate in terms of $|V_{ub}|$ 
rely on an operator 
product expansion (OPE) in inverse powers of the $b$ 
quark mass \cite{ope} and thus  
depend on the choice of renormalization scale and include non-perturbative
parameters.
Experimentally, the principal difficulty is the separation of $B \rightarrow
X_u e\nu$ decays from the dominant $B\rightarrow X_c e\nu$ decays.   
Selection criteria applied to achieve this separation generally make it 
difficult to translate  the observed rate to the full decay rate.  

In this paper we present a measurement of the inclusive electron spectrum
for charmless semileptonic $B$ decays in the momentum range of
$2.3 - 2.6$~\gevc\ as measured in the \FourS\ rest frame.
In the rest frame of the $B$ meson, the
kinematic endpoint of the electron spectrum for the dominant 
$B \rightarrow X_c e \nu$ decays is $\sim2.3 \gevc$ and 
$\sim 2.6 \gevc$ for $B \rightarrow X_u e \nu$ decays. 
In the rest frame of the \FourS, the $B$ mesons have a momentum 
of $\sim 0.3\gevc$ and thus the electron spectrum is convoluted by a
spread of $\pm 0.2$\gevc, extending the 
endpoint to higher energies. Nevertheless, a narrow interval of
about $300\,\mathrm{MeV/c}$ remains that is dominated by electrons
from $B \rightarrow X_u e \nu$ transitions. This interval covers 
approximately 10\% of the total electron spectrum for charmless
semileptonic $B$ decays.  The 
extrapolation from the limited momentum range near the endpoint
to the full spectrum is a very difficult task because the OPE breaks 
down in this part of phase space. 

This analysis is based on the same method as previous measurements 
of the lepton spectrum near the endpoint 
\cite{argus1, argus2} (ARGUS), \cite{cleo1, cleo2} (CLEO).

\section{Detector and Data Sample}
This analysis is based on data recorded in 1999-2000 with the 
\babar\ detector at the \pep2 energy asymmetric $e^+e^-$ collider
at the Stanford Linear Accelerator Center. The data sample 
corresponds to an integrated luminosity 
of 20.6~$\mathrm{fb}^{-1}$ that was collected at the \FourS\ resonance 
(\textrm{ON}),
plus an additional sample of 2.6~$\mathrm{fb}^{-1}$ that was recorded 
about 40~MeV below the \FourS peak (\textrm{OFF}). 

The relative normalization of the \textrm{ON} and \textrm{OFF}-peak data 
is $7.87\pm0.01\pm0.04$. This factor has been derived from luminosity 
measurements, which are based on the QED cross section for 
$e^+e^-\to \mu^+\mu^-$ 
production, corrected for the energy dependence of the cross section. 
The systematic error on the relative normalization is estimated to be 
$\sim 0.5\%$.  This error accounts for small variations in the detector 
response and is significantly smaller than the systematic error of 
$\sim 1.6\%$ on the absolute $\mu^+\mu^-$ cross section measurement.

The \babar\ detector has been described in detail elsewhere \cite{BBRD}.
The most important components for this study are the charged particle 
tracking system, consisting of a five-layer 
silicon detector and a 40-layer drift chamber, and the electromagnetic 
calorimeter assembled from 
6580 CsI(Tl) crystals. Electron candidates are selected on the 
basis of the ratio of the energy detected in the calorimeter 
to the track momentum, the calorimeter shower shape, 
the energy loss in the drift chamber, and the angle reconstructed 
in the ring imaging Cherenkov detector.  

The electron identification efficiency and the probabilities to misidentify 
a pion, kaon, or proton as an electron have been measured with clean 
samples of tracks 
that were selected from the data. This experimental information is 
introduced into the Monte Carlo simulation to improve the agreement with the 
data. Tracking efficiencies and resolution have been studied. A comparison 
with the simulation has revealed small differences, which have been taken 
into account. No significant impact of non-Gaussian resolution tails 
has been found in the endpoint region.

The Monte Carlo simulation of $B \rightarrow X_u e \nu$ events is based
on the ISGW2 model~\cite{ISGW2}.  In the current version, the hadrons $X_u$ are
represented by single particles or resonances with masses up to 
$1.5$~\gevcc and non-resonant contributions are not included.  
For $B \rightarrow X_c e\nu$ transitions three models are
employed to simulate different decay modes. 
The decay to $D^* e \nu$ is modeled following a form factor based 
parameterization of HQET~\cite{hqet}; for decays to $D e \nu$ and 
higher mass charm meson states the ISWG2 model is used.
The non-resonant decays to $D^{(*)}\pi e \nu$  are modeled 
according to a prescription by Goity and Roberts~\cite{gr}.

\section{Data Analysis}

\subsection{Event Selection}

For this analysis, electron candidates are selected in the
momentum range from 1.5 to 3.5~\gevc in the \FourS\ rest frame. The solid angle
is restricted by the electromagnetic calorimeter coverage, defined 
by the laboratory polar angle  range of 
$-0.72<\cos \theta_{\mathrm{lab}}<0.92$. In this momentum and angular range, 
the efficiency for identifying an electron has been measured to be 
$\epsilon_{PID} = 0.91 \pm 0.02$, while the average hadron 
misidentification probability is less than $0.2\%$. 
The selected electron sample is dominated by electrons from semileptonic 
decays of $B$ and $D$ mesons, non-resonant $q\bar{q}$ production and QED 
processes.  In addition, photon conversions and Dalitz decays contribute 
background at low momenta and $J/\psi$ decays contribute at higher momenta. 
Furthermore, 
there are sizable contributions from hadrons misidentified as electrons. 
To suppress low-multiplicity QED processes, including 
$\tau^+\tau^-$ pairs, the number of charged tracks per event is required 
to be greater than three. This background and non-resonant hadronic 
events are further suppressed by a restriction on the ratio of 
Fox-Wolfram moments $H_2/H_0<0.4$ \cite{foxw}. 
In semileptonic decays, the neutrino carries a substantial momentum. 
In events in which the only undetected particle is this neutrino, 
its direction can be inferred from the missing momentum in the event, 
defined as the difference between the net momentum of the two colliding beam 
particles and the vector sum of all detected particles, 
charged and neutral. Therefore, 
the selection of these decays can be greatly enhanced by requiring that the
measured missing momentum 
exceed 1~\gevc\ and point into the detector fiducial volume.
Furthermore, since the $B$ mesons are produced almost at rest and the high
momentum electron and neutrino point in nearly opposite directions, we require
that the angle between the electron candidate and the missing momentum 
be greater
than $\pi/2$. Candidate electrons are rejected if, when paired with an
opposite-sign electron, the invariant mass of the pair is consistent with
$J/\psi$ mass, $3.05 < M_{e^+e^-} < 3.15$~\gevcc.

The detection efficiencies are estimated using Monte Carlo simulation. 
For the selection criteria described above, 
the detection efficiency for charmless semileptonic decays in the electron 
momentum interval of $1.5 - 2.7$~\gevc\ ranges from $\sim 0.4$ 
to $\sim 0.25$.

\subsection{Background Subtraction} 

\begin{figure}[!htb]
\begin{center}
\resizebox{0.9\textwidth}{0.9\textwidth}{
\includegraphics{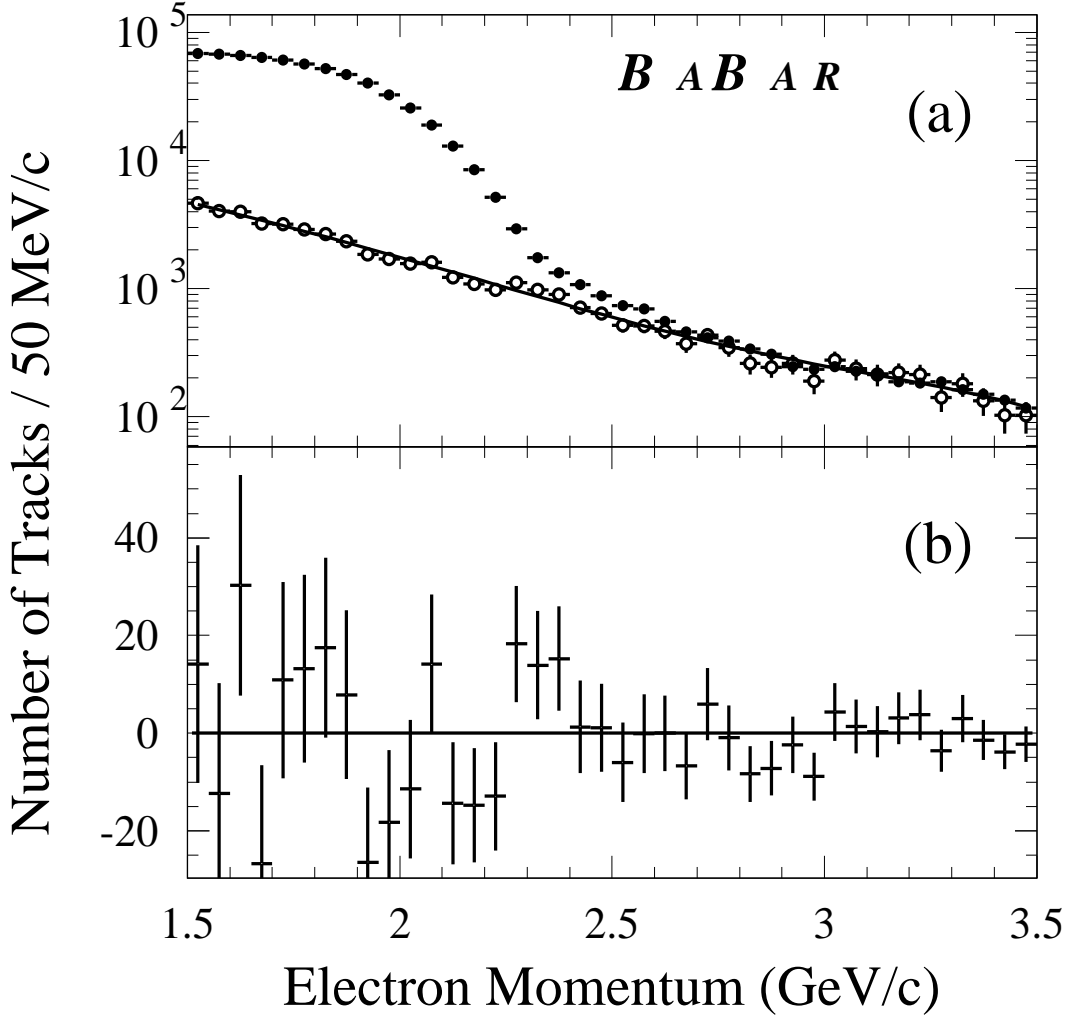}
}
\caption{
 Electron momentum spectrum in the $\FourS$ rest frame: 
(a) \textrm{ON}-peak data (solid circles), 
\textrm{OFF}-peak data scaled by the \textrm{ON/OFF}-ratio (open circles), 
(b) \textrm{OFF}-peak data (unscaled) after the subtraction of the 
fitted continuum. The line show the result of a fit to the continuum 
spectrum (using both \textrm{ON}- and \textrm{OFF}-peak data) 
in the interval 
$p_{\mathit{cms}} = 1.5 - 3.5$~\gevc with a 4-th degree 
Chebyshev polynomial.
}
\label{e1_spectrum}
\end{center}
\end{figure}

The raw spectrum of the highest momentum electron in events selected
by the criteria described above is shown in
Figure~\ref{e1_spectrum}a, separately for data recorded \textrm{ON} 
and \textrm{OFF} resonance. 
For the \textrm{OFF}-resonance data, the momenta 
are scaled to compensate for the 0.4\% difference in the c.m.s. energies 
of the two data samples. 

Also shown in Figure~\ref{e1_spectrum}a is a fit to approximate the 
non-resonant background contribution.  For the purpose of this fit 
the normalized \textrm{OFF}-peak data in the interval  of 
$p_{\mathit{cms}} = 1.5 - 3.5$~\gevc are combined with 
the \textrm{ON}-peak data 
in the interval  of $p_{\mathit{cms}} = 2.7 - 3.5$~\gevc
above the kinematic limit for $B\to X_u e \nu$ decays. 
The $\chi^2/\mbox{dof}$ for a fit with a 4-th degree Chebyshev 
polynomial is 51/51. 
Figure~\ref{e1_spectrum}b shows the residuals of the 
fitted function from the unscaled \textrm{OFF}-peak data. 
The fit significantly reduces the statistical uncertainty 
in the estimate of the continuum background that is to be subtracted.  

\begin{figure}[!htb]
\begin{center}
\resizebox{0.9\textwidth}{0.9\textwidth}{
\includegraphics{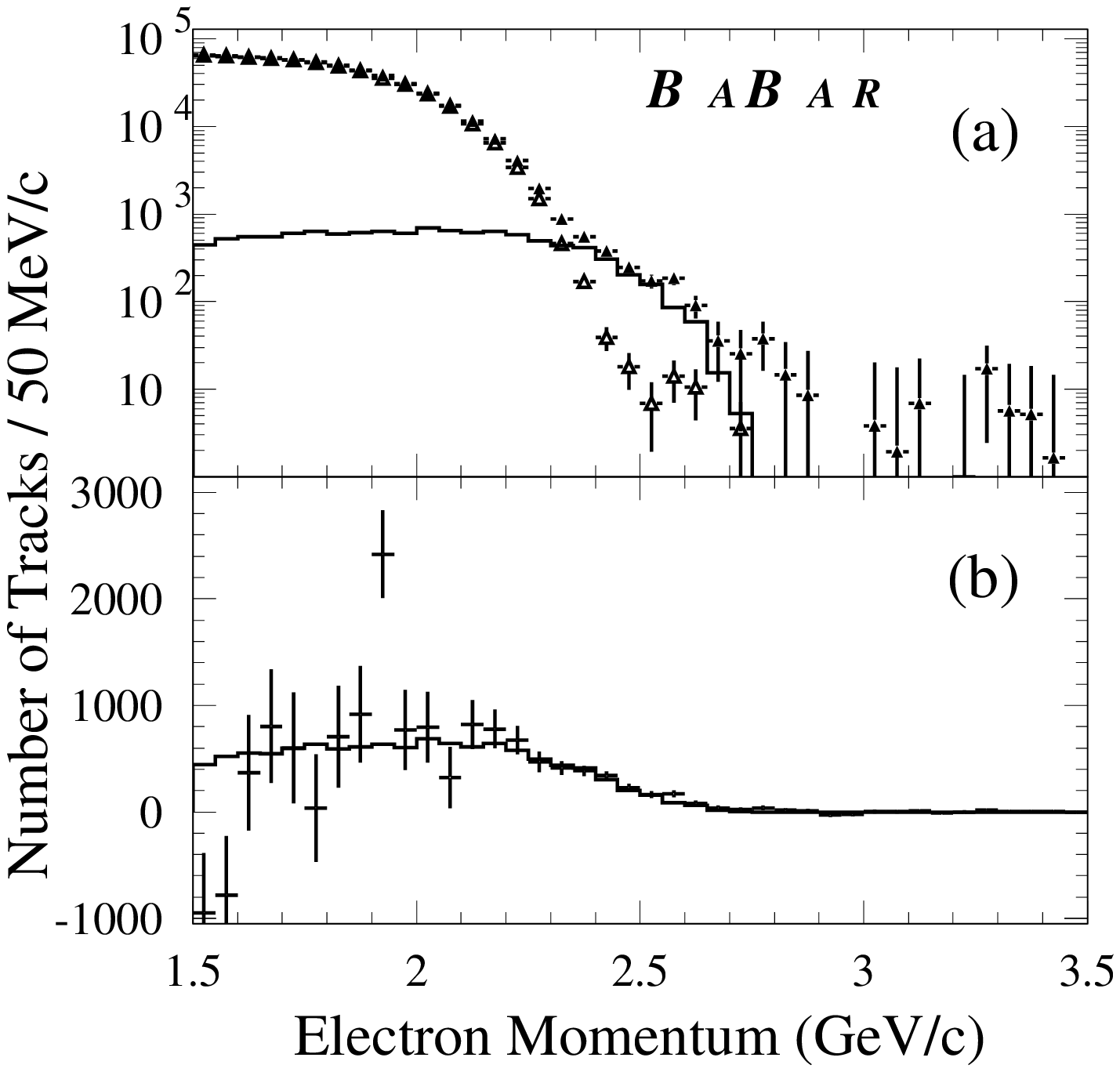}
}
\caption{
The electron momentum spectrum in the $\FourS$ rest frame: 
(a) \textrm{ON}-peak data after continuum subtraction (solid triangles), and
MC predicted background from $B\bar{B}$ events ($B \nrightarrow X_u e \nu$) 
(open triangles). 
(b) \textrm{ON} peak data after subtraction of continuum and 
Monte Carlo predicted ($B \nrightarrow X_u e \nu$) backgrounds 
(data points with statistical errors). 
For comparison, the histograms show the expected signal spectrum from 
$B \to X_u e \nu$ decays.
}
\label{e2_spectrum}
\end{center}
\end{figure}

The result of the subtraction of the fitted continuum background is 
shown in Figure~\ref{e2_spectrum}a. 
Also shown are the Monte Carlo predictions of the 
expected signal from $B \to X_u e \nu$ decays and background 
contributions from all 
other processes, $B \nrightarrow  X_u e \nu$. In this simulation, 
the $B \to X_u e \nu$ branching ratio is assumed to be $1 \cdot 10^{-3}$ 
and the relative normalization of data and Monte Carlo simulation is set 
by the number of electrons in the momentum interval between 1.5 and 2.3~\gevc. 
The result of the subtraction of all backgrounds is shown 
in Figure~\ref{e2_spectrum}b.

The number of selected events, split into three momentum intervals,
together with the estimated background contributions are
listed in Table~\ref{t:evt_num}. The intervals are chosen
to emphasize differences in the background composition: 
the interval 2.0 and 2.3~\gevc\ is dominated by semileptonic 
$B \to X_c e \nu$ decays, with a small contribution from hadron 
misidentification, 
the $2.3 - 2.6$~\gevc\ interval is mostly  populated by  $B\to X_u
e \nu$ decays and continuum background, while the
$2.7 - 3.0$~\gevc\ interval contains almost exclusively continuum 
background.

\begin{table}[!htb]
\caption{
The number of signal and background events and signal selection 
efficiencies $\epsilon(B\to X_u e \nu)$
for three different intervals in the electron momentum. 
$N_\mathrm{ON}$ refers to the number of selected electrons recorded on the 
$\FourS$ resonance, $N_{\mathrm{OFF}}$ is the fitted number 
of continuum background events. 
The $N(B \to X_c e \nu)$ and $N(B \to X_c \to e)$
refer to the Monte Carlo estimate of the remaining semileptonic background,
from prompt and secondary decays (including $\psi(2S)$), respectively. 
$N(B\to J/\psi \to e^+e^-) $ is an
estimate of $J/\psi$ background remaining after the mass cut. 
$N(B\to X_c)$ is a background from hadronic $B$ decays to charm. 
$N(B\to X_u e \nu)$ is the resulting number of signal electrons.
}
{\small
\begin{center}
\begin{tabular}{llll} \hline \hline
$Momentum~p_e$~(\gevc) 
 &$2.0 - 2.3$      & $2.3 - 2.6$ & $2.7 - 3.0$ \\ \hline
$N_{\mathrm{ON}}$  & $74,140 \pm 272$ & $6,455 \pm 80$ & $1,932 \pm 44$ \\
$N_{\mathrm{OFF}}$ & $7,749  \pm 165$ & $4,051 \pm 93$& $1,903 \pm 37$ \\
$N(B \to X_c e \nu)$ & 
$61,158 \pm 470$   & $470 \pm 41$ & $0 $ \\
$N(B \to J/\psi \to e^+e^-) $ & $666 \pm 49$ & $128 \pm 22$ & $0$ \\
$N(B \to X_c \to e )$ &
$338 \pm 35$   & $18 \pm 8$ & $0 $ \\ 
$N(B \to X_c)$ - mis-ID & $373 \pm 37$  & $92 \pm 18$ & $4 \pm 4$ \\ 
\hline
$N(B\to X_u e \nu)$ 
                   & $3,857 \pm 572$    & $1,696 \pm 133$& $25 \pm 57$ \\
\hline
$\epsilon(B\to X_u e \nu)(\%)$ 
                   & $33.9 \pm 1.0$     & $27.7 \pm 1.3$ & $28.6 \pm 22.9$\\ 
\hline \hline
\end{tabular}
\end{center}
}
\label{t:evt_num}
\end{table}

\subsection{Determination of $B\to X_u e \nu$ Branching Ratio}

For a given interval in the electron momentum, the inclusive
partial branching ratio is calculated according to 
\begin{equation}
\Delta \textrm{\BR} = \frac{N_{\mathrm{ON}}-N_{\mathrm{OFF}}-N_{B 
\nrightarrow X_u e \nu}}
{2\epsilon N_{B\bar{B}}}(1+\delta_{\mathit{rad}}).
\end{equation}
 
\begin{figure}[!htb]
\begin{center}
\resizebox{0.9\textwidth}{0.9\textwidth}{
\includegraphics{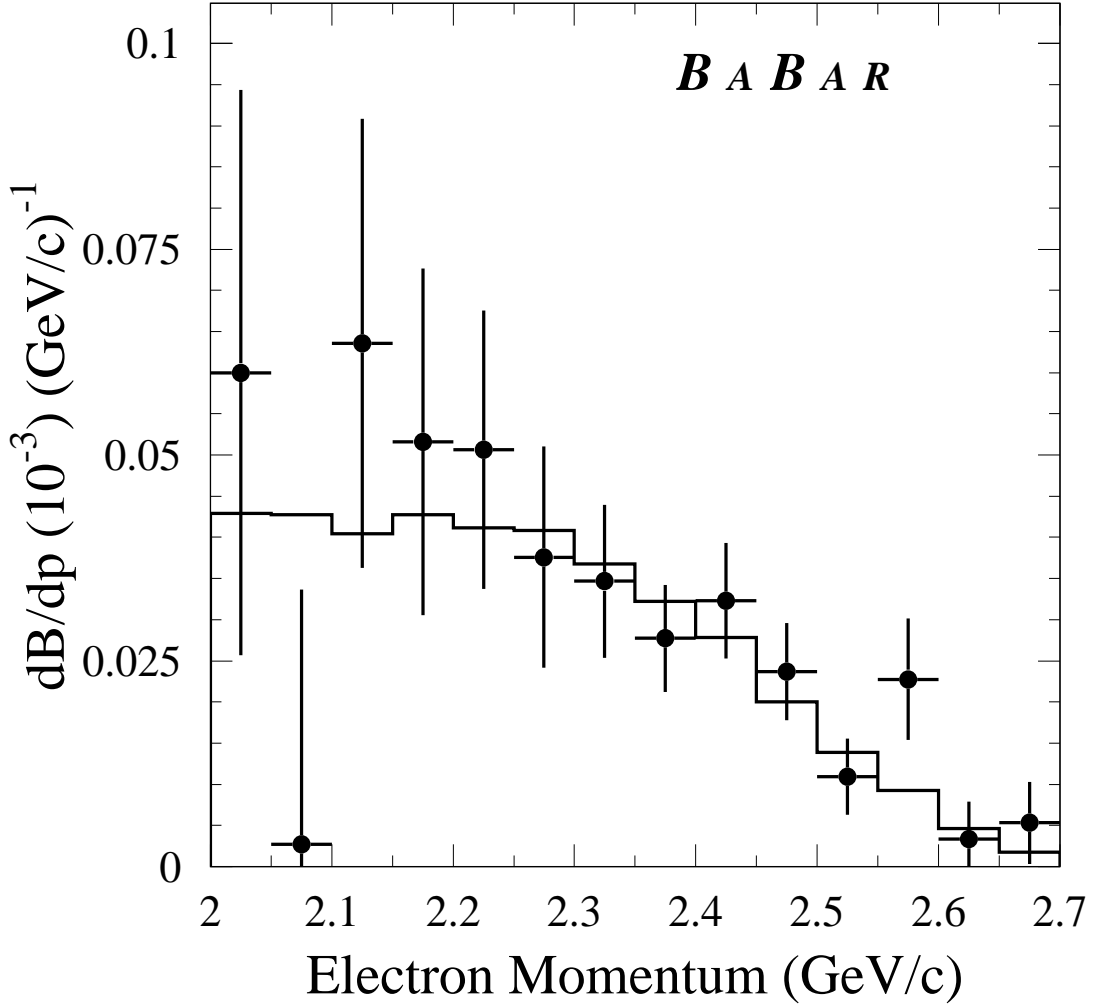}
}
\caption{
The branching ratio $\textrm{\BR}(B \rightarrow X_u e \nu)$ 
as a function of 
the electron momentum in the \FourS rest frame. The data (statistical
errors only) are compared to a
prediction (solid line) based on the ISGW2 model assuming a total 
inclusive branching 
ratio  of $1 \cdot 10^{-3}$ for $B \rightarrow X_u e\nu$ decays 
with $X_u$ masses up to $1.5$~\gevcc.
The spectrum is corrected for final-state radiation and bremsstrahlung.
}
\label{fsp1t}
\end{center}
\end{figure}

\noindent 
Here $N_{\mathrm{ON}}$  refers to the number of 
electrons detected \textrm{ON}-peak and $N_{\mathrm{OFF}}$ refers to 
the fitted continuum background in a specified momentum interval,
$N_{{B} \nrightarrow X_u e \nu}$ is the 
background from $B\bar{B}$ events derived from Monte Carlo simulation 
normalized to the total number of electrons in the momentum range 
$1.5 - 2.3$~\gevc, 
$\epsilon$ is the total efficiency for detecting a signal electron from 
$B \rightarrow X_u e\nu$ decays (including bremsstrahlung in detector 
material), and $\delta_{\mathit{rad}}$ accounts for the 
distortion of the electron spectrum due to final-state radiation. 
This last is a momentum dependent correction, derived from the 
Monte Carlo 
simulation, amounting to $\sim 11\%$ in the range $2.3 - 2.6$~\gevc.
As the overall normalization the total number of produced $B\bar{B}$ events 
is used, $N_{B\bar{B}} = (22,630 \pm 19 \pm 362) \cdot 10^3$.
The differential branching ratio as a function of the electron momentum 
in the \FourS\ rest frame is shown in 
Figure~\ref{fsp1t}. Integrated over the interval 
from 2.3 to 2.6~\gevc, the partial branching ratio is 
(statistical error only):
\begin{equation}
\Delta \textrm{\BR}(B \rightarrow X_u e \nu) = (0.152\pm 0.014)\cdot 
10^{-3}.
\end{equation}  

\subsection{Systematic Errors}

The following sources of systematic errors have been considered: 
\begin{itemize}
\item
the Monte Carlo simulation of the signal events and the impact on the 
detection efficiency, 
\item
continuum background subtraction, 
\item
the subtraction of all other backgrounds based on Monte Carlo simulation
of the production process and the detector simulation,
\item
the uncertainty of the $B$ meson momentum spectrum in the \FourS rest 
frame.
\end{itemize}

The systematic error introduced by the efficiency estimation for the 
signal events is caused by 
uncertainties in simulation of event selection (4\%), the track 
reconstruction (1\%), and the electron identification efficiencies (2\%).
The error on the selection efficiency depends not only on the accuracy 
of the detector model, but also on the adequacy of the ISGW2 model on 
which the event generation is based. The
latter error has been estimated from the impact of the variation 
(within their current uncertainties) of individual branching 
ratios for $B$ meson decays and variation of the event selection cuts.
The total systematic error on the efficiency amounts to $5\%$.

An error in the continuum background subtraction can be introduced by
the choice of the fitting function and the fitting procedure. 
The impact of the choice of the function has been estimated by trying 
different functional forms, such as an exponential 
function and Chebyshev polynomials.
Possible differences in continuum spectra in \textrm{ON}-peak and 
\textrm{OFF}-peak data have been assessed by including or excluding 
\textrm{ON}-peak data above $B\to X_u e \nu$ kinematic limit in the fit. 
The observed variations in the number of continuum events 
leads us to attribute a 
2\% systematic uncertainty on the number of continuum events. This is
equivalent to $5\%$ systematic error in the number of signal events.

The $b \to c$ background consists of electrons and misidentified hadrons 
(see Table~\ref{t:evt_num}). The dominant contribution are electrons from 
semileptonic $B\to X_c e \nu$ decays. There is also a much smaller 
contribution of secondary electrons from $B\to X_c\to X_s e \nu$ 
and $B\to \psi(2S) \to ee$ decays, but very few with momenta above 2.3~\gevc. 
The error in the estimate of prompt lepton  background arises primarily 
from the 
uncertainty in the decay dynamics and the branching fractions of the 
individual decay modes. The impact of these uncertainties (4\%) is 
estimated by variation of these branching ratios within their current 
uncertainties. 
In addition, the uncertainty in the electron identification (2\%) has 
been included. The systematic error (28\%) on the background from 
misidentified 
hadrons arises from both the uncertainty in the spectrum of charged 
hadrons above 2.3~\gevc\ and the overall uncertainty on the 
mis-identification probabilities for pions and kaons. 
In addition, there is a small correction for electrons from $J/\psi$ 
decays that are not vetoed by the mass requirement; the uncertainty 
in this correction is 20\%.
Background from photon conversions and Dalitz decays have been analyzed and 
found to be negligible above 1.5~\gevc. 
The total systematic error on the Monte Carlo based $b\to c$  background 
subtraction is estimated to be 6\%, translating to a relative error of 
3\% on the number of signal events. 

Since the \FourS mass is so close to the threshold for $B\bar{B}$ 
production, the $B$ meson momentum and thereby the endpoint of the electron 
spectrum from $B\to X_c e\nu$ decays is highly sensitive to the total 
energy of the colliding beams. Thus variations of the colliding beam energy 
introduces a systematic error in the $B\to X_c e\nu$
background subtraction. These variations are included in the Monte Carlo 
simulations, and the impact of the difference between data and Monte Carlo 
is studied using a sample of fully reconstructed $B$ mesons. 
The comparison of the measured and simulated momentum distributions, 
currently limited by statistics, shows that the systematic error from this 
effect does not exceed 5\%.

In summary, Table \ref{t:syst} contains list of systematic errors for the 
momentum range from 2.3 to 2.6~\gevc; the total systematic error on 
the partial branching ratio measurement for this momentum range 
is $\sim 9\%$.

\begin{table}[!htb]
\caption{Systematic error on the $\Delta \textrm{\BR}$ for 
the momentum range from 2.3 to 2.6~\gevc. 
All numbers are given in percent. 
}
{\small
\begin{center}
\begin{tabular}{lcc} \hline \hline
contribution & $\delta x/x$ & 
$\delta(\Delta\textrm{\BR})$/$\Delta\textrm{\BR}$  
\\ \hline 
efficiency $\epsilon(B\to X_u e \nu)$ & 5 & 5 \\
continuum subtraction & 2 & 5  \\
background $B \to X_c e \nu$ & 4 & 1 \\
background $B \to J/\psi \to e^+e^-$ & 20 & 2 \\
background $B \to X_c$ - mis-ID & 28  & 2 \\
$B$ movement & 5 & 5 \\ 
$N_{B\bar{B}}$ & 1.6 & 1.6 \\
radiative corrections & 10 & 1 \\
\hline
total & -- & 9 \\
\hline \hline
\end{tabular}
\end{center}
}
\label{t:syst}
\end{table}

\section{Conclusion}

The result of this analysis, the fully corrected differential branching 
ratio as a function of the electron momentum in the \FourS\ rest frame, 
is presented in 
Figure~\ref{fsp1t}. Integrating over the interval from 2.3 to 2.6~\gevc 
results in the partial branching ratio 
\begin{equation}
\Delta \textrm{\BR} (B\rightarrow X_u e\nu)=(0.152\pm 0.014\pm 0.014)\cdot 10^{-3}.
\end{equation}

\noindent 
This result agrees very well with the measurement by the CLEO 
collaboration \cite{cleo2}.
The measured electron spectrum in the endpoint region is well reproduced 
by the ISGW2 model.

\section{Extraction of $|V_{ub}|$ based on CLEO Measurements}

To determine the charmless semileptonic branching fraction 
$\textrm{\BR} (B\rightarrow X_u e\nu)$ from the partial branching fraction 
$\Delta \textrm{\BR}(\Delta p)$, one needs to 
know the fraction $f_u(\Delta p)$ of the spectrum that falls into 
the momentum interval $\Delta p$. Heavy Quark theory describes 
the Fermi-motion of the quarks inside the meson in terms of 
a shape function that depends on non-perturbative QCD. 
To leading order, the same shape function describes all 
$b \ra q \ell \nu$ transitions (here $q$ represents any light quark).

The CLEO collaboration \cite{CLEO-result} has recently used the measurement 
of the inclusive photon spectrum from $b\to s\gamma$ transitions to derive 
the parameters describing the shape function and to calculate the lepton 
momentum spectrum for 
$B\rightarrow X_u e\nu$ transition.  They quote a value of 
$f_u(\Delta p) = 0.074 \pm 0.014 \pm 0.009$ for the interval $\Delta p$
from 2.3 to 2.6~\gevc. 

Relying on the CLEO measurement, the result presented here 
translates into a total branching ratio of
\begin{equation}
\textrm{\BR}(B\rightarrow X_u e\nu)=
(2.05 \pm 0.27_{exp} \pm 0.46_{f_u})\cdot 10^{-3}.
\end{equation}

Based on studies developed independently by two groups \cite{V1}-\cite{V6}, 
we adopt the following relationship for the extraction of 
$|V_{ub}|$ from the total branching fraction \cite{CERN01050} 
(revised most recently at the 2002 CKM workshop) 
\begin{equation}
|V_{ub}| = 0.00445 \left( \frac{\textrm{\BR}(B \to X_u l \nu)}{0.002} 
\frac{1.55 \mbox{ps}} {\tau_b} \right)^{1/2} 
(1.0 \pm 0.020 \pm 0.052), 
\end{equation} 
\noindent
where the first error arises from the uncertainty in the OPE expansion,  
and the second from the uncertainty in the $b$ quark mass, 
assuming $m_b(1 \gev)=4.58 \pm 0.09 \gev$. 
Based on this expression and our measurement of the branching 
fraction, we find 
\begin{equation}
|V_{ub}|= 
(4.43 \pm 0.29_{exp} \pm 0.25_{OPE} \pm 0.50_{f_u} \pm 0.35_{s\gamma}) \cdot 10^{-3}.
\end{equation}
Here the first error is the measurement uncertainty from the 
combined statistical and systematic error, and the second refers to the
uncertainty on the extraction of $|V_{ub}|$ from the branching ratio 
as stated above. 
The remaining errors are taken from the CLEO analysis: 
the third refers to the determination of $f_u$  
and the fourth represents an estimate of the validity of the assumption 
that the shape function can be extracted from the $b \ra s \gamma$ spectrum.
The uncertainty due to the theoretical assumption of quark-hadron 
duality remains unquantifiable.

The \babar\ collaboration is planning to develop this and other methods 
for extracting the total branching ratio and $|V_{ub}|$ in the near future.

\section{Acknowledgments}

We are grateful for the 
extraordinary contributions of our \pep2\ colleagues in
achieving the excellent luminosity and machine conditions
that have made this work possible.
The success of this project also relies critically on the 
expertise and dedication of the computing organizations that 
support \babar.
The collaborating institutions wish to thank 
SLAC for its support and the kind hospitality extended to them. 
This work is supported by the
US Department of Energy
and National Science Foundation, the
Natural Sciences and Engineering Research Council (Canada),
Institute of High Energy Physics (China), the
Commissariat \`a l'Energie Atomique and
Institut National de Physique Nucl\'eaire et de Physique des Particules
(France), the
Bundesministerium f\"ur Bildung und Forschung and
Deutsche Forschungsgemeinschaft
(Germany), the
Istituto Nazionale di Fisica Nucleare (Italy),
the Research Council of Norway, the
Ministry of Science and Technology of the Russian Federation, and the
Particle Physics and Astronomy Research Council (United Kingdom). 
Individuals have received support from 
the A. P. Sloan Foundation, 
the Research Corporation,
and the Alexander von Humboldt Foundation.


\begin{thebibliography}{99}

\bibitem{ope}
J.\ Chay, H.\ Georgi, and B.\ Grinstein, Phys. Lett. 
{\bf B247}, 399 (1990).

\bibitem{argus1}   
The ARGUS Collaboration, H.\ Albrecht {\em et al}., Phys. Lett. {\bf B234}, 409 (1990). 
%
\bibitem{argus2}  
The ARGUS Collaboration, H.\ Albrecht {\em et al}., Phys. Lett. {\bf B255}, 297 (1991).
%
\bibitem{cleo1}   
The CLEO Collaboration, R.\ Fulton {\em et al}., Phys. Rev. Lett. {\bf 64}, 16 (1990). 
%
\bibitem{cleo2}   
The CLEO Collaboration, F.\ Bartelt {\em et al}., Phys. Rev. Lett. {\bf 71}, 4111 (1993). 
%
\bibitem{BBRD}    
The \babar\ Collaboration, B.\ Aubert {\em et al}., Nucl. Instrum. Methods. 
{\bf A479}, 1 (2002).

\bibitem{ISGW2}
N.\ Isgur, D.\ Scora, B.\ Grinstein, and M.B.\ Wise, Phys. Rev. 
{\bf D39}, 799 (1989); 
D.\ Scora, N.\ Isgur, Phys. Rev. {\bf D52}, 2783 (1995).
%
\bibitem{hqet}
I.I.\ Bigi, M.\ Shifman, and N.G.\ Uraltsev, Annu. Rev. Nucl. Part. Sci. 
{\bf 47}, 591 (1997).
%
\bibitem{gr}
J.L.\ Goity and W.\ Roberts, Phys. Rev. {\bf D51}, 3459 (1995).
%
\bibitem{foxw}
G.C.\ Fox and S.\ Wolfram, Phys. Rev. Lett. {\bf 41}, 1581 (1978). 
%
\bibitem{CLEO-result}
The CLEO Collaboration, A.\ Bornheim {\em et al}., hep-ex/0202019 (2002).
%
\bibitem{V1}      
N.\ Uraltsev, Int. J. Mod. Phys. {\bf A11}, 515 (1996).
%
\bibitem{V2}       
I.I.\ Bigi, M.\ Shifman, and N.\ Uraltsev, TPI-MINN-97/02-T, 
UMN-TH-1528-97, UND-HEP-97-BIG01, hep-ph/9703290 (1997).
%
\bibitem{V3}
N.\ Uraltsev, Int. J. Mod. Phys. {\bf A14}, 4641 (1999).
%
\bibitem{V4}
I.I.\ Bigi, UMN-HEP-BIG-99-05, hep-ph/9907270 (1999).
%
\bibitem{V5}       
A.H.\ Hoang, Z.\ Ligeti, and A.V.\ Manohar, Phys. Rev. Lett. 
{\bf 82}, 277 (1999).
%
\bibitem{V6}       
A.H.\ Hoang, Z.\ Ligeti, and A.V.\ Manohar, Phys. Rev. 
{\bf D59}, 074017 (1999).
%
\bibitem{CERN01050}
Combined results on b-hadron production rates and decay properties,
CERN-EP/2001-050 (2001).

\end{thebibliography}
\end{document}